\documentclass[final,5p,times,twocolumn]{elsarticle}

\usepackage{graphicx}
\usepackage{dcolumn}
\usepackage{bm}
\usepackage{amsmath}
\usepackage{amssymb}

\usepackage{lineno,hyperref}
\modulolinenumbers[5]

\journal{Journal of Quantitative Spectroscopy and Radiative Transfer}

\begin{document}

\begin{frontmatter}

\title{The branching ratio of intercombination $A^1\Sigma^+\sim b^3\Pi\to a^3\Sigma^+/X^1\Sigma^+$ transitions in the RbCs molecule: measurements and calculations\tnoteref{label1}}

\tnotetext[label1]{This paper is dedicated to the memory of our collegue and friend Pupyshev Vladimir Ivanovich}


\author{V. Krumins}
\author{A. Kruzins}
\author{M. Tamanis}
\author{R. Ferber\fnref{ferbermail}}
\fntext[ferbermail]{ferber@latnet.lv}
\address{Laser Center, Faculty pf Physics, Mathematics and Optometry, University of Latvia, 19 Rainis blvd, Riga LV-1586, Latvia}

\author{A. Pashov\fnref{pashovmail}}
\fntext[pashovmail]{pashov@phys.uni-sofia.bg}
\address{Faculty of Physics, Sofia University, 5 James Bourchier blvd., 1164 Sofia, Bulgaria}

\author{A. V. Oleynichenko}
\author{A. Zaitsevskii}
\address{Petersburg Nuclear Physics Institute named by B.P. Konstantinov of National Research Center ``Kurchatov Institute'', 188300 Gatchina, Leningrad District, Russia}
\address{Department of Chemistry, Lomonosov Moscow State University, 119991, Moscow, Leninskie gory 1/3, Russia}

\author{E. A. Pazyuk}
\author{A. V. Stolyarov\fnref{stolyarovmail}}
\fntext[stolyarovmail]{avstol@phys.chem.msu.ru}
\address{Department of Chemistry, Lomonosov Moscow State University, 119991, Moscow, Leninskie gory 1/3, Russia}

\begin{abstract}
We observed the $A^1\Sigma^+\sim b^3\Pi\to a^3\Sigma^+/X^1\Sigma^+$ laser-induced fluorescence (LIF) spectra of the RbCs molecule excited from the ground $X^1\Sigma^+$ state by the Ti:Sapphire laser. The LIF radiation from the common perturbed levels of the singlet-triplet $A\sim b$ complex was recorded by the Fourier-transform (FT) spectrometer with the instrumental resolution of 0.03~cm$^{-1}$. The relative intensity distribution in the rotationally resolved $A\sim b\to a^3\Sigma^+(v_a)/X^1\Sigma^+(v_X)$ progressions was measured, and their branching ratio was found to be about of 1$\div$5$ \times$10$^{-4}$ in the bound region of the $a^3\Sigma^+$ and $X^1\Sigma^+$ states. The experiment was complemented with the scalar and full relativistic calculations of the $A/b - X/a$ transition dipole moments (TDMs) as functions of internuclear distance. The relative systematic error in the resulting \emph{ab initio} TDM functions evaluated for the strong $A - X$ transition was estimated as few percent in the energy region, where the experimental LIF intensities are relevant. The relative spectral sensitivity of the FT registration system, operated with the InGaAs diode detector and CaF beam-splitter, was calibrated in the range $[6~500,12~000]$~cm$^{-1}$ by a comparison of experimental intensities in the long $A\sim b\to X(v_X)$ LIF progressions of the K$_2$ and KCs molecules with their theoretical counterparts evaluated using the \emph{ab initio} $A - X$ TDMs. Both experimental and theoretical transition probabilities can be employed to improve the stimulated Raman adiabatic passage process, $a\to A\sim b \to X$, which is exploited for a laser assembling of ultracold RbCs molecules.
\end{abstract}


\end{frontmatter}

%\linenumbers

\section{Introduction}\label{ruvin-intro}

\begin{figure}%[t!]
\center
\includegraphics[scale=0.4]{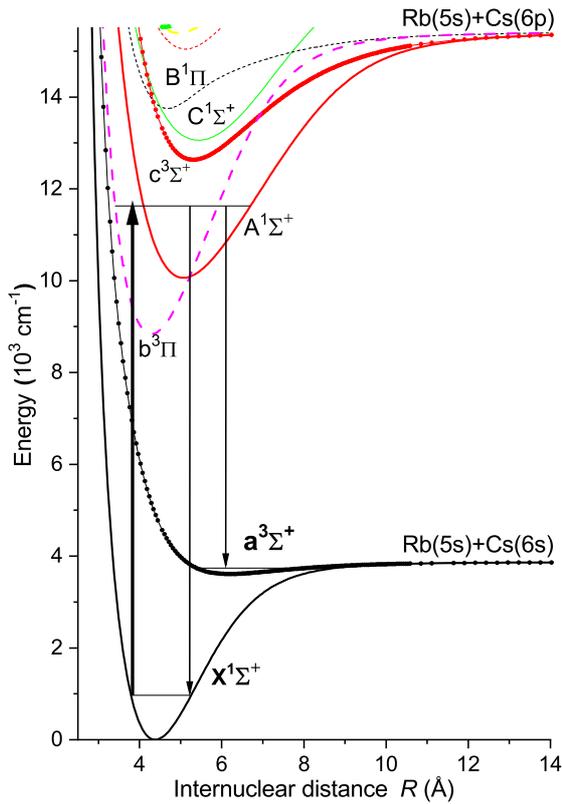}
\caption{Term scheme of the ground and lowest excited electronic states of the RbCs molecule according to the scalar-relativistic calculation~\cite{Allouche2000}.}\label{Fig_PEC}
\end{figure}

In alkali metal diatomic molecules containing the heavy Rb or Cs atoms the first excited $A^1\Sigma^+$ and $b^3\Pi$ electronic states are completely mixed due to the strong spin-orbit interaction leading to the formation of the singlet-triplet $A^1\Sigma^+\sim b^3\Pi$ complex ($A\sim b$ complex for short). The comprehensive studies of the $A\sim b$ complex are of interest for several reasons. First, the strong singlet-triplet mixing makes it possible to reach the higher-lying triplet states manifold from the ground singlet $X^1\Sigma^+$ state through a wide window of rovibronic levels belonging to the $A\sim b$ complex, see for instance Refs.~\cite{Li1983, Lazarov2001, Manaa2002}. Second, the $A\sim b$ complex is exploited as an intermediate state in the optical paths for transforming ultracold molecular species formed (by magnetoassociation or photoassociation) in the vaguely bonded $a^3\Sigma^+$ state levels to the absolute ground $X^1\Sigma^+(v_X=0,J_X=0)$ state, see reviews~\cite{Ulmanis2012, Pazyuk2015}. Specifically, in~ \cite{Debatin2011, Nagerl2014, Molony2014} the ultracold RbCs molecules production was performed by coherent two-steps stimulated Raman adiabatic passage (STIRAP) process~\cite{Vitanov2017}. The first laser of STIRAP pumps the $A\sim b$ levels from weakly bound Feshbach levels of the initial $a^3\Sigma^+$ state while the second laser dumps the excited $A\sim b$ levels to the ground $X^1\Sigma^+$ state. So, the detailed knowledge of probabilities of both $a\leftrightarrow A\sim b$ and $ A\sim b \leftrightarrow X$ transition, see Fig.~\ref{Fig_PEC},~Ref.~\cite{Allouche2000}, is of practical interest to improve the STIRAP efficiency. However, to our best knowledge, in spite of the favorable conditions provided by the singlet-triplet nature of the perturbed $A\sim b$ levels the spontaneous radiative $A\sim b\to a^3\Sigma^+$ transitions in alkali diatomics have not been observed yet, except NaK molecule~\cite{Masters1990}, in which the $A\sim b\to a^3\Sigma^+(v_a)$ LIF intensity distributions were measured allowing to determine the empirical $b^3\Pi-a^3\Sigma^+$ transition dipole moment (TDM). The reason for this situation is the extremely small probabilities of the ``triplet'' $A\sim b\to a$ transitions.

The comprehensive spectroscopic analysis of the strongly perturbed rovibronic levels of the $A\sim b$ complex for RbCs has been performed in Refs.\cite{Docenko2010, Kruzins2014}, while the accurate empirical potential energy curves (PECs) were constructed for the isolated ground $X^1\Sigma^+$ and $a^3\Sigma^+$ states in Ref.~\cite{Docenko2011}. In \cite{Docenko2011}, data on the triplet $a^3\Sigma^+$ state were obtained by using the routes from higher excited states, namely by the $B^1\Pi\sim c^3\Sigma^+\sim b^3\Pi \to a^3\Sigma^+$ and $(4)^1\Sigma^+\to a^3\Sigma^+$ laser induced fluorescence (LIF) spectra.

The radiative and predissociative lifetimes of vibrational levels belonging to the RbCs $A\sim b$ complex were systematically evaluated in Ref.~\cite{Londono2011}. The required $A - X$ and $b - a$ TDM  functions were obtained in the framework of simple electronic structure model~\cite{Aymar2005, Aymar2012} based on full configuration interaction treatment of the two-valence-electron problem defined by the large-core two-component pseudopotentials of alkali atoms and the $l$-dependent core-polarization potentials~\cite{Fuentealba1983} (2$e$-CPP). The relativistic configuration-interaction valence-bond method has been used in Ref.~\cite{Kotochigova2005} to estimate the interatomic potentials and TDMs between the ground and first excited states of RbCs as well.

The aim of the present study is to perform experimental and theoretical investigation of the $A\sim b\to X/a$ transition probabilities based on relative intensity measurements and calculations of the vibrational LIF progressions of the RbCs molecule. We aim to determine the vibronic branching ratio of very weak ``triplet'' $A\sim b\to a^3\Sigma^+(v_a)$ and rather strong ``singlet'' $A\sim b\to X^1\Sigma^+(v_X)$ transitions to the bound part of $X^1\Sigma^+$ and $a^3\Sigma^+$ states.

As far as experiment is concerned, the most challenging issue is to combine the ultra-high spectral resolution needed to select the particular rovibronic LIF progression of the dense spectra, which has to be provided by the high-resolution FT spectrometer, with high sensitivity necessary to detect very weak spectra. What is more, calibration of the spectral sensitivity of the FTS registration system in a wide spectral range is indispensably required. In present study it has been accomplished one in the unconventional way, namely: by means of a direct comparison of the experimental relative intensity distribution in the long $A\sim b\to X^1\Sigma^+(v_X)$ LIF progressions of ''test'' KCs and K$_2$ molecules with their theoretical counterparts evaluated using the spin-allowed $A^1\Sigma^+-X^1\Sigma^+$ \emph{ab initio} TDMs as a function of internuclear distance $R$.

%The \emph{ab initio} evaluated electronic TDMs are unambiguously responsible for the main part of uncertainty in the radiative properties predicted for RbCs molecules as well as K$_2$ and KCs.
To assess a reliability of the estimated rovibronic transition probabilities both scalar and full relativistic $A/b - X/a$ TDM functions, which are obtained in the framework of high level electronic structure calculations, were systematically exploited. It should be also noted that the present measurements and calculations were greatly facilitated by the spectroscopically accurate deperturbation model previously constructed for the $A\sim b$ complex of the RbCs~\cite{Kruzins2014}, KCs~\cite{Kruzins2013}, and K$_2$~\cite{Manaa2002} molecules, as well as by the highly accurate ground state PECs of these molecules, see Refs.~\cite{Docenko2011, Ferber2013, Pashov2008}.

\section{Measurements}\label{maris-experiment}
\subsection{Experimental setup}\label{maris-setup}

In the present experiment RbCs molecules were excited by the Ti:Sapphire laser radiation from the ground $X^1\Sigma^+$ state into a particular rovibronic level of the $A^1\Sigma^+\sim b^3\Pi$ complex. The corresponding LIF transitions to the ground singlet $X^1\Sigma^+$ and the lowest triplet $a^3\Sigma^+$ states were recorded, see Fig.~\ref{Fig_PEC}. RbCs molecules were produced at about 300$^\circ$C in a linear heat-pipe filled with Rb and Cs metals. Laser beam was sent into the heat-pipe through the pierced mirror and the backward LIF was recorded using the Fourier Transform (FT) spectrometer IFS125-HR (Bruker) with spectral resolution 0.03~cm$^{-1}$ in the spectral range from 12~000~cm$^{-1}$ to 6~000~cm$^{-1}$. Laser power typically was about 300-400~mW. The exploited laser frequencies were selected within the range 10~780 -- 11~010~cm$^{-1}$. A standard InGaAs room temperature diode with sensitivity maximum near 6~500~cm$^{-1}$ was used to detect the LIF spectra.

\subsection{Assignment of the $A\sim b\to X/a$ LIF spectra}\label{maris-analysis}

We aimed to excite the rovibronic levels of the $A\sim b$ complex exhibiting a substantial mixing of $A^1\Sigma^+$ and $b^3\Pi$ states. The particular levels for excitation were selected from the rovibronic term values database of the $A\sim b$ complex calculated with the deperturbed molecular parameters from Ref.~\cite{Kruzins2014}. As proved in Ref.~\cite{Alps2017} the database reproduces the rovibronic term values of the $A\sim b$ complex with accuracy typically better than 0.01~cm$^{-1}$, hence the required laser frequency could be predicted with appropriate accuracy. The database, which can be found in the Supplement of Ref.~\cite{Alps2017}, also contains the mixing coefficients of the states treated.

In order to observe very weak LIF transitions to the triplet $a^3\Sigma^+$ state the FT-LIF experiment was realized in several steps. First, we recorded and analyzed the full rather strong $A\sim b\to X^1\Sigma^+$ LIF progressions excited by a predicted laser frequency $\nu_L$ to be sure the observed spectra involved a particular LIF progression from the targeted upper state of the $A\sim b$ complex and that the $\nu_L$ excites this level in a strongest absorption transition with respect to ground state vibrational levels $v_X$. The unambiguous rotational and vibrational assignment of the RbCs $A\sim b\to X^1\Sigma^+(v_X)$ progressions was performed by means of a highly accurate empirical PEC available for the $X^1\Sigma^+$ state~\cite{Docenko2011}. No optical filters were used during this step. To avoid detector's saturation by scattered laser light, the lowest signal amplification was implemented. At the next step we cut off the most part of the $A\sim b\to X^1\Sigma^+(v_X)$ spectra by an optical long-pass filter FEL1100 placed in the observation path, with transmission upper limit at about 9~100~cm$^{-1}$, and recorded the LIF spectra with the increased amplification. We searched for the $A\sim b\to a^3\Sigma^+(v_a)$ transitions in the spectral range from 7~000 to 8~000~cm$^{-1}$. If succeeded, at the final step the filter FEL1200 was used, which cut off LIF above 8~300~cm$^{-1}$; the spectrum was recorded with the highest amplification and increased acquisition time.

\begin{figure}%[t!]
\center
\includegraphics[scale=0.4]{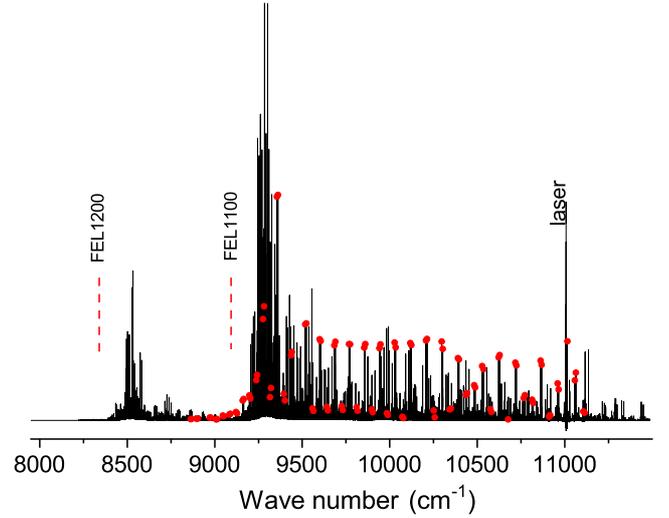}
\caption{The LIF spectrum recorded at lowest signal amplification using the excitation laser frequency $\nu_L$ = 11~007.865~cm$^{-1}$. No optical filter is used. The spectrum contains 62 LIF $A\sim b\to X$ progressions assigned to the RbCs, Rb$_2$, and Cs$_2$ molecules. The red points on top of lines mark the $P$,$R$-progression from the targeted level $J^{\prime}=97$, $E_{A\sim b}$=11~291.672~cm$^{-1}$ of the $^{85}$Rb$^{133}$Cs isotopomer excited in the $A\sim b\leftarrow X(v_X=2;J_X = 98)$ transition. Vertical dashed lines mark approximate transmission upper limits for the long-pass edge  FEL1100 and FEL1200 filters.}\label{Fig_LIF1}
\end{figure}

\begin{figure}%[t!]
\center
\includegraphics[scale=0.4]{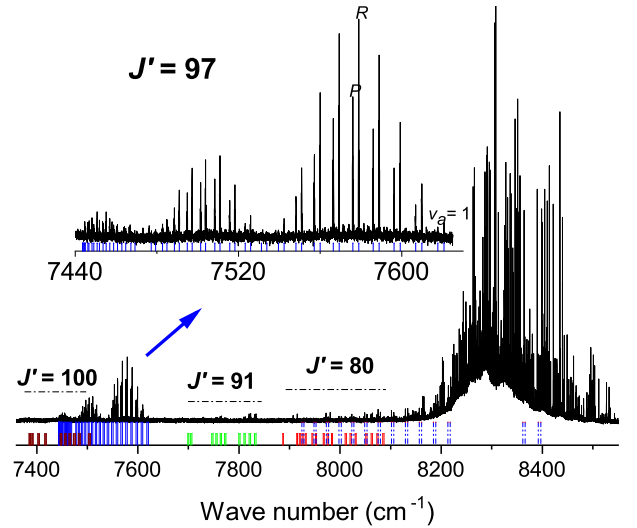}
\caption{Fragment of the same LIF spectrum as presented in Fig.~\ref{Fig_LIF1} while recorded with the FEL1200 filter and increased by about 40 times amplification. Blue (long) bars below the spectrum mark a progression to the triplet $a^3\Sigma^+$ state from the targeted $J^{\prime}$=97 level (see zoomed in spectrum in the inset). Well separated groups of short bars (wine, red and green) mark progressions from the accidentally excited levels with $J^{\prime}$=80, 100 ($^{85}$Rb$^{133}$Cs) and 91 ($^{87}$Rb$^{133}$Cs), respectively. Dashed blue bars mark the very end of the assigned progression from $J^{\prime}$=97 to the ground $X^1\Sigma^+$ state.}\label{Fig_LIF2}
\end{figure}

This procedure is illustrated by Figs.~\ref{Fig_LIF1} and~\ref{Fig_LIF2}. The spectrum presented in Fig.~\ref{Fig_LIF1} was recorded without filter at the lowest amplification. Along with the particular progression from the targeted level $A\sim b (J^{\prime} = 97, E_{A\sim b} = 11~291.672$~cm$^{-1})\to X(v_X, J_X = 96, 98)$ of the $^{85}$Rb$^{133}$Cs isotopomer it contains many other accidentally excited RbCs, Rb$_2$, and Cs$_2$ progressions, most of them being very weak. After increasing amplification by about 40 times and cutting off the LIF with the FEL1200 filter, the LIF transitions to the $a^3\Sigma^+$ state could be observed as shown in Fig.~\ref{Fig_LIF2}. The strongest doublet progression to the $a^3\Sigma^+$ state in the range from 7~440 to 7~620~cm$^{-1}$ is marked by long blue bars; it originates from the targeted level of the $A\sim b$ complex with $J^{\prime}$=97 and $E_{A\sim b}$=11~291.672~cm$^{-1}$. In the inset of Fig.~\ref{Fig_LIF2} it is seen that in $P$,$R$-doublets the $R$-branch is stronger than the $P$-branch by about 30\%. Some other much weaker fragmentary $A\sim b\to a(v_a)$ progressions starting from accidentally excited levels of the $A\sim b$ complex are also seen in this spectrum.

\begin{figure}%[t!]
\center
\includegraphics[scale=0.6]{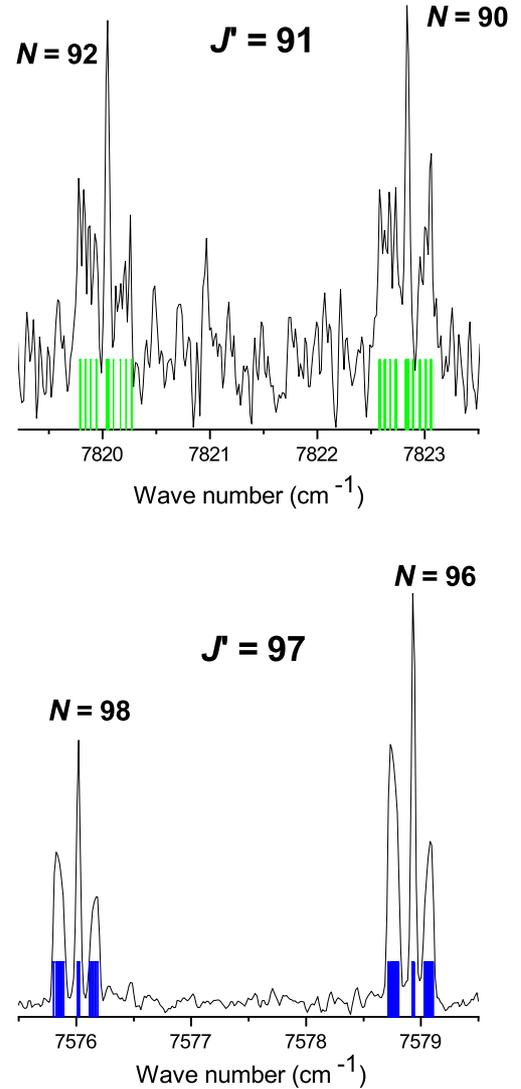}
\caption{Hyperfine structure of $^{85}$Rb$^{133}$Cs (lower panel) and $^{87}$Rb$^{133}$Cs (upper panel) observed in doublet $P$,$R$-progressions to the $a^3\Sigma^+$ state with the rotational quantum number $N_a = J^{\prime} \pm 1$ from the levels of the $A\sim b$ complex with $J^{\prime}=97$ and $J^{\prime}=91$, respectively. Vertical bars mark positions of the HFS components estimated in Section~\ref{Asen-hfs}.}\label{Fig_HFS}
\end{figure}

Due to the large hyperfine structure (HFS) in the triplet $a^3\Sigma^+$ state each line in the $A\sim b\to a$ progressions is split into several groups as shown in Fig.~\ref{Fig_HFS}. For the $^{85}$Rb$^{133}$Cs isotopomer (see lower panel) the strong central group is accompanied by two weaker side groups. Analysis of the lines with good signal-to-noise ratio (SNR) for a number of progressions proved that the splitting and the intensity ratio of HFS groups remain unchanged for different vibrational bands.

The HFS pattern of the $A\sim b\to a$ lines must in principle depend also on the hyperfine splitting of the upper state levels belonging to the $A\sim b$ complex. However, all of the observed transitions show the same HFS, independent of the upper state quantum number $J^{\prime}$ and energy. As is known, see for instance~\cite{Matsubara1993}, the HFS of the excited state $b^3\Pi$ in NaRb is smaller than the Doppler width of the excitation transition, so that all HFS components of the upper state are sufficiently populated. It is interesting to note that in the studies of the $a^3\Sigma^+$ state in a number of heteronuclear alkali diatomics exploiting $B^1\Pi\sim c^3\Sigma^+\to a^3\Sigma^+$ transitions~\cite{Docenko2011,Pashov2005,Staanum2007}, a similar HFS stability of LIF lines was observed, in spite of the fact that the hyperfine splitting in the $c^3\Sigma^+$ state is expected to be much larger then in the $b^3\Pi_0$ state, see e.g. Ref~\cite{Matsubara1993}. Only in the case of KRb~\cite{Pashov2007}, the changes in HFS of the LIF lines changes at scanning the laser around the center of the excitation frequency were observed. Apparently in this case the hyperfine splitting of the upper state levels exceeds the Doppler broadening, and, hence, some of the HFS components could not be populated and the corresponding transitions to the $a^3\Sigma^{+}$ state levels are missing.

\begin{figure}%[t!]
\center
\includegraphics[scale=0.33]{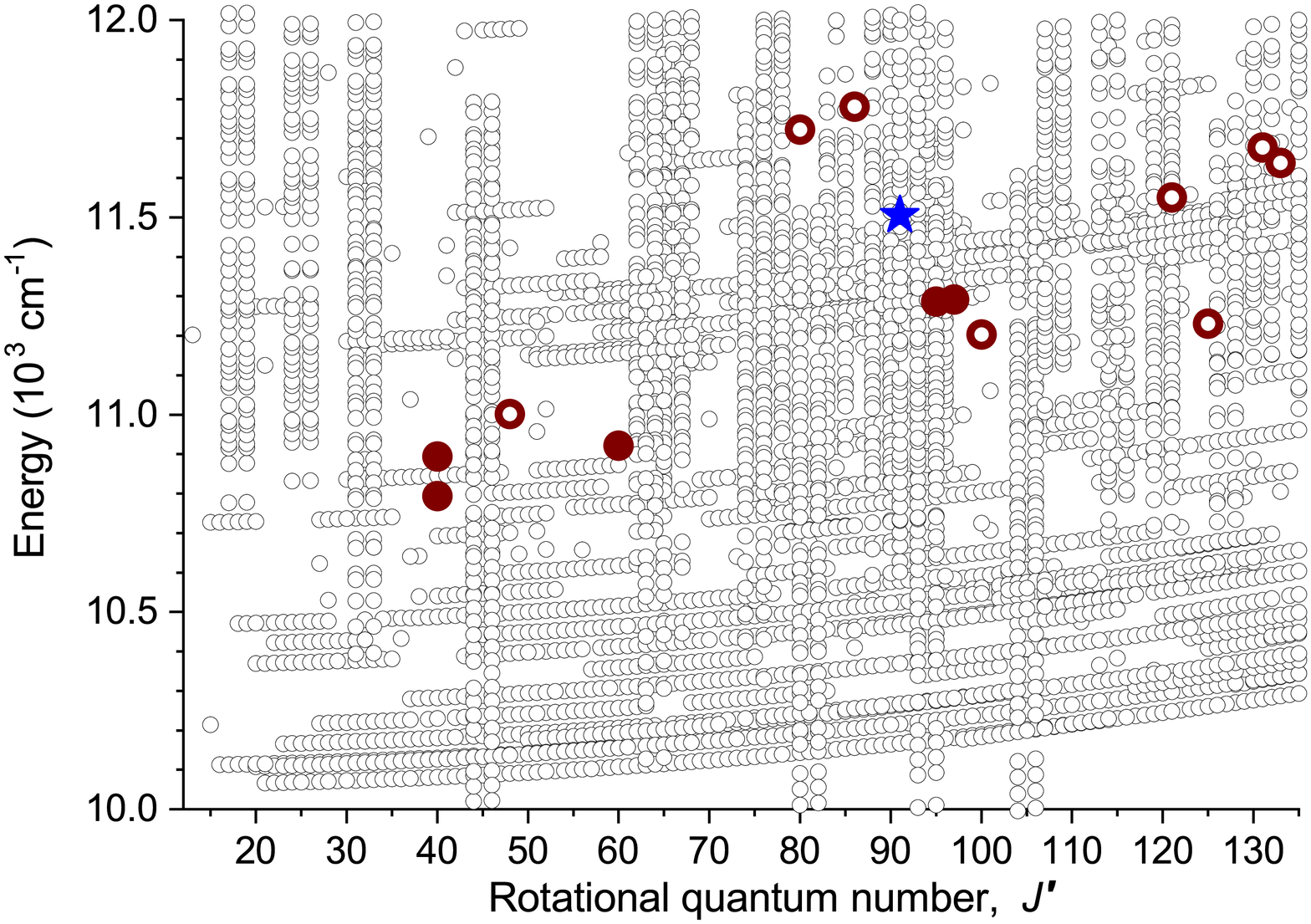}
\caption{The rovibronic levels of the $A^1\Sigma^+\sim b^3\Pi$ complex of the $^{85}$Rb$^{133}$Cs isotopomer, from which the $A\sim b\to X^1\Sigma^+/a^3\Sigma^+$ LIF progressions were observed. Solid red circles mark  the targeted levels; red open circles - the accidentally excited levels; blue star -the $^{87}$Rb$^{133}$Cs isotopomer; open grey circles belong to a fragment of the overall $A\sim b$ experimental term value data set borrowed from Ref.~\cite{Kruzins2014}.} \label{Fig_term}
\end{figure}

In the present experiment overall 14 $A\sim b\to a(v_a)$ LIF progressions were recorded, among them 5 from the targeted levels and 9 from accidentally excited levels; the latter includes one progression belonging to the $^{87}$Rb$^{133}$Cs isotopomer, see Fig.~\ref{Fig_HFS} (upper panel). Fig.~\ref{Fig_term} overviews their distribution over the upper state rotational quantum number $J^{\prime}$. For better visibility the figure also contains a fragment of the overall data set of $A\sim b$ experimental term values from Ref.~\cite{Kruzins2014}. It can be seen that the present measurements cover the energy region from about 10~800 till 11~800~cm$^{-1}$, which is by ca 800~cm$^{-1}$ above the bottom of the singlet $A^1\Sigma^+$-state, see Figs.~\ref{Fig_PEC} and \ref{Fig_term}.

\subsection{Intensities of the $A\sim b\to X/a$ LIF progressions}\label{maris-intensity}

Intensity measurements in the recorded LIF spectra have been performed in order to determine relative intensity distributions within rovibrational structure of both $A\sim b\to X^1\Sigma^+(v_X)$ and $A\sim b\to a^3\Sigma^+(v_a)$ LIF progressions originated from a common upper level of the $A\sim b$ complex, as well as to determine the branching ratios for the observed transitions to triplet and singlet states.

Relative intensity distribution within a particular progression was measured for $P$- and $R$-branches separately, see Figs.~\ref{Fig_LIF1} and~\ref{Fig_LIF2}. The line intensity was determined by a peak value of line profile. In the measurements of relative intensity distributions within triplet $A\sim b\to a^3\Sigma^+$ progressions the peak value of the central group $I_{cg}$, see Fig.~\ref{Fig_HFS},  was used as well since the central component was stronger than the side components for all recorded transition to the triplet state, hence it provided the best SNR value.

The spectra recorded at the same excitation frequency with different filters and amplification (a) without filter with FEL1100, and with FEL1200 filters, were step-by-step matched by comparing the intensities of common lines of a particular progressions under study. Only the lines with sufficient SNR and far enough from the filters transmission limits were used for this purpose.

In order to compare calculated and measured intensities of transitions to the $a^3\Sigma^+$ state, the experimental total line intensity was determined in the following way. Since in the $a^3\Sigma^+$ state all HFS transitions to a particular rotational level of the $^{85}$Rb$^{133}$Cs isotopomer form three groups of partially overlapping lines, see Fig.~\ref{Fig_HFS}, the contribution of each group into the total line  intensity was obtained by integration of the area below line profile of each group. Only the lines having SNR above 10 were used for this purpose. Averaged over many lines, normalized ratios of side wings intensities to the central component's intensity were determined as 1.3 (left) / 1.0 (central) / 0.8 (right), see Fig.~\ref{Fig_HFS}, which yield a correction value 3.1. Additional correction was applied, since we observed that the central group is slightly broadened if compared with the width of lines to the singlet state, obviously due to the HFS. The full width at half maximum of the central group and of the line to the singlet state was about 0.045~cm$^{-1}$ and 0.035~cm$^{-1}$ respectively. The ratio [Intensity(peak value)/Area below line profiles] for the central group and for transitions to the singlet state, averaged over many measurements, showed that the peak value of the central group should be multiplied by 1.14. Hence, the total intensity of lines to the triplet state was $I_{A\sim b\to a}=I_{cg}$ (peak value)$\times$(3.54$\pm$0.25). The branching ratios for LIF transitions to the $a^3\Sigma^+$ and $X^1\Sigma^+$ states from a common level of the $A\sim b$ complex were obtained, according to Eq.(\ref{Ratio}) from Sec.~\ref{Andrey-CC}, by summation over all observed lines in the respective bound-bound transitions to the triplet and singlet states.

\subsection{Spectral sensitivity of the FT spectrometer detection system}\label{maris-sensitivity}

\begin{figure}%[t!]
\center
\includegraphics[scale=0.4]{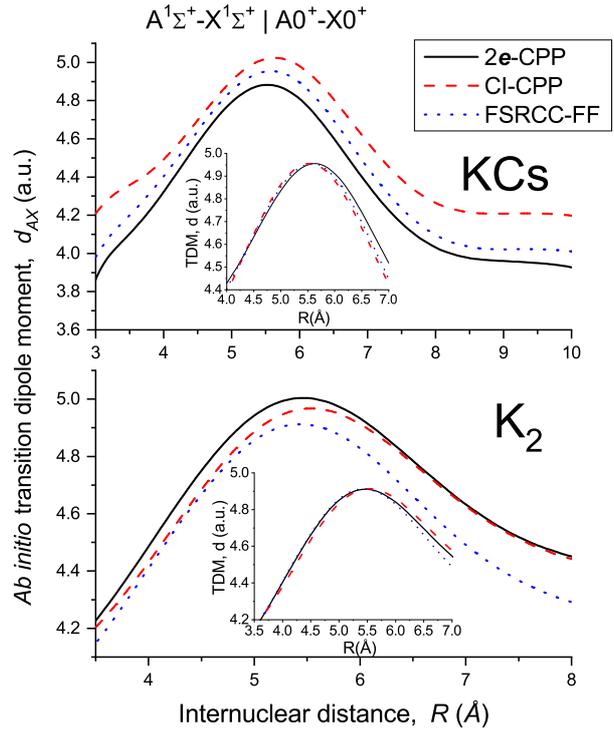}
\caption{The spin-allowed $A^1\Sigma^+-X^1\Sigma^+|A_{0^+}-X_{0^+}$ TDM functions $d^{ab}_{AX}(R)$ obtained for K$_2$ and KCs molecules in the framework of various \emph{ab initio} approaches: 2$e$-CPP~\cite{Aymar2012}, CI-CPP, and FSRCC-FF (see details in Sec.~\ref{zay-relativism}). The insets demonstrate the coincidence of the uniformly scaled TDM functions initially obtained by different methods.}\label{Fig_TDM_K2KCs}
\end{figure}

In determination of correct intensity distributions and branching ratios of electronic transitions in the long LIF progressions spanning over broad spectral range a calibration of spectral sensitivity of the detection system is crucially important. The relative spectral sensitivity $S(\nu)$ of the our FT spectrometer operated with InGaAs diode detector at room temperature and CaF beam-splitter, was determined in a wide spectral range $\nu \in [6~500, 12~000]$~cm$^{-1}$ in a following way. First, in the range $\nu \in [8~250, 12~000]$~cm$^{-1}$ the required $S(\nu)$ function was obtained as a ratio of experimental relative line intensities $I^{Expt}$ previously measured for a large amount of the $A\sim b\to X(v_X)$ LIF progressions of both K$_2$ and KCs molecules with their theoretical counterparts $I^{Calc}$
\begin{eqnarray}\label{Sn}
S(\nu)=\frac{I^{Expt}_{A\sim b\to X}}{I^{Calc}_{A\sim b\to X}}.
\end{eqnarray}
The required $I^{Calc}$-values were evaluated according to Eq.(\ref{Itensinglet}) in Sec.~\ref{Andrey-CC} by means of the highly accurate deperturbation parameters available for the $A\sim b$ complex of both molecules~\cite{Manaa2002,Kruzins2013} and corresponding spin-allowed $A^1\Sigma^+-X^1\Sigma^+$ TDM functions $d^{ab}_{AX}$ calculated by various \emph{ab initio} methods in Sec.~\ref{zay-relativism}. In particular, the scalar-relativistic CI-CPP functions (see \ref{CICPP}) were employed for calibration purpose.

As is easy seen from Fig.~\ref{Fig_TDM_K2KCs} the \emph{ab initio} $d^{ab}_{AX}$ functions obtained in the framework of scalar- and full-relativistic approximations are very close to each other. Furthermore, the simple uniform scaling of the original 2$e$-CPP, CI-CPP and FSRCC-FF TDMs yields almost undistinguishable curves since the residual divergence of the scaled TDM functions does not exceed 1-2\% at intermediate internuclear distances (see insets in Fig.~\ref{Fig_TDM_K2KCs}). Therefore the main part of a systematic error in the sensitivity curve $S(\nu)$ is determined by the uncertainty in the experimental LIF intensities, which can reach 10-15\% for weak lines.

\begin{figure}%[t!]
\center
\includegraphics[scale=0.44]{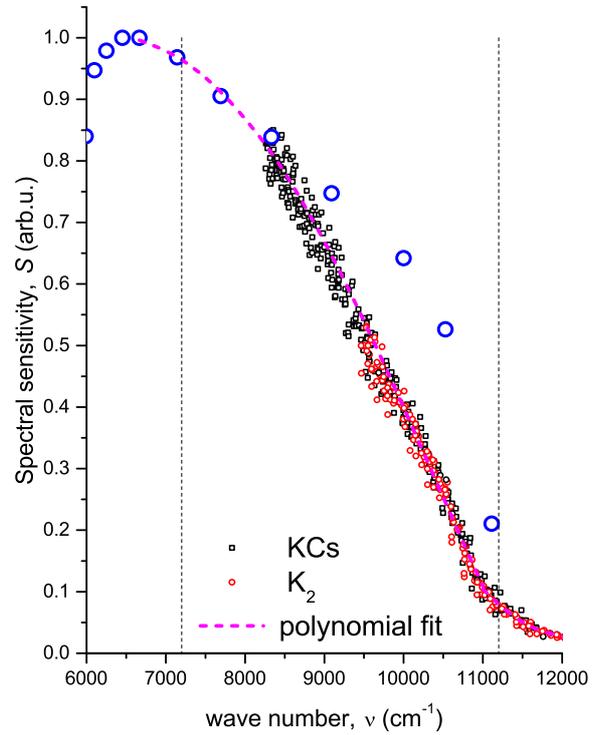}
\caption{The relative spectral sensitivity curve $S(\nu)$ obtained for the present FT detecting system operated with InGaAs diode and CaF beam splitter. The piece-wise $S(\nu)$-function was constructed by smooth matching the S$(\nu)$ curve from the K$_2$ and KCs intensity data, obtained according to Eq.~(\ref{Sn}), with the InGaAs diode sensitivity curve (large open circles) from \cite{Hamamatsu}. Dashed line denotes the fitted polynomial function $S(\nu)$. Thin vertical lines mark the spectral range covered in the present RbCs experiment.}\label{Fig_sensitivity}
\end{figure}

It should be noted that the LIF progressions of the K$_2$ dimer were accidentally excited and recorded during our previous studies of the $A\sim b$ complex in KCs and KRb molecules~\cite{Kruzins2013,Alps2016}. The Fig.~\ref{Fig_sensitivity} clearly demonstrates that the LIF results of the KCs and K$_2$ molecules used for the FTS calibration are well overlapped in a rather wide spectral range $\nu \in [9~500, 12~000]$~cm$^{-1}$.

To proceed to the lower frequency range where the present $A\sim b\to a^3\Sigma^+(v_a)$ progressions of RbCs are located, the ``semi-empirical'' function $S(\nu)$ obtained according to Eq.~(\ref{Sn}) was smoothly matched near the region of $\nu$~=~8~300~cm$^{-1}$ with the standard InGaAs detector sensitivity curve borrowed from Ref.~\cite{Hamamatsu}. In the spectral range from 6~500 to 8~500~cm$^{-1}$ the $S(\nu)$-dependence is slightly (by about 25\%) decreasing, see Fig.~\ref{Fig_sensitivity}. The combined point-wise $S(\nu)$ curve in the entire range [6~500, 12~000]~cm$^{-1}$ was approximated by a polynomial function, which was used to normalize the present intensity measurements in the RbCs molecule.

\section{Calculations}

\subsection{Intensity distributions and branching ratios of the $A\sim b\to X/a$ transitions}\label{Andrey-CC}

The pronounced SO-coupling effect in RbCs and KCs molecules leads to strong mutual perturbations of the close-lying $A^1\Sigma^+$ and $b^3\Pi$ states (see Fig.~\ref{Fig_PEC}). Therefore, a rigorous deperturbation treatment is indispensably needed to represent the energies of the $A\sim b$ complex and the $A\sim b\to X/a$ transition probabilities at an experimental level of confidence~ \cite{Pazyuk2015,Pazyuk2019}. The required non-adiabatic eigenvalues $E_{A\sim b}$ and corresponding multi-component eigenfunctions $\phi_i$ ($i\in[A^1\Sigma^+;b^3\Pi_{0^+}; b^3\Pi_1;b^3\Pi_2]$) of the rotational $J^{\prime}$-level of the $A\sim b$ complex were obtained numerically by solving four coupled-channel (CC) radial equations~\cite{Yurchenko2016} for all three molecules involved (see Appendix A).
The eigenvalues and eigenfunctions for the isolated ground $X^1\Sigma^+$ and $a^3\Sigma^+$ states in RbCs (see Fig.~\ref{Fig_PEC}) were obtained by solving the single-channel radial equation with the corresponding empirical (adiabatic) PECs~\cite{Docenko2011}. Hyperfine coupling of $X^1\Sigma^+$ and $a^3\Sigma^+$ states taking place near the common dissociation limit was completely ignored during the present study.

The similar spectroscopic model was applied for the calculation, in accordance to Eq.(\ref{Itensinglet}), of the relative intensities in the particular rovibrational $A\sim b\to X(v_X)$ LIF progressions measured previously for the K$_2$ and KCs molecules (see Sec.~\ref{maris-sensitivity}). The required molecular parameters such as the empirical interatomic PECs of the interacting $A$ and $b$ states, the refined SO $A\sim b$ coupling functions and the highly accurate ground $X$-state PECs were borrowed from Refs.~\cite{Manaa2002,  Kruzins2013, Ferber2013, Pashov2008}.

The relative intensity distribution in the vibrational LIF progressions $I$ and branching ratios $\mathbb{R}$ of $A\sim b\to X/a$ transitions coming from a common rovibronic levels of the $A\sim b$ complex to rovibrational levels of the lower-lying $X^1\Sigma^+$ and $a^3\Sigma^+$ states can be evaluated as
\begin{eqnarray}\label{Itensinglet}
I_{A\sim b\to X}(J^{\prime},v_X,J_X) \sim \nu^4_{A\sim b-X}|\langle \phi_A^{J^{\prime}}|d^{ab}_{AX}|\chi_X^{J_X}\rangle|^2,
\end{eqnarray}
\begin{eqnarray}\label{Itentriplet}
I_{A\sim b\to a}(J^{\prime},v_a,N_a) \sim \nu^4_{A\sim b-a}|\langle \phi_{b0^+}^{J^{\prime}}|d^{ab}_{ba}|\chi_a^{N_a}\rangle|^2,
\end{eqnarray}
\begin{eqnarray}\label{Ratio}
\mathbb{R}_{a/X} = \frac{\sum_{v_a}\sum_{N_a}I_{A\sim b\to a}}{\sum_{v_X}\sum_{J_X}I_{A\sim b\to X}},
\end{eqnarray}
where $\nu_{A\sim b-X/a}$ are the wavenumbers of the $A\sim b\to X/a$ intercombination transitions, $|\nu_X^{J_X}\rangle$ and $|\nu_a^{N_a}\rangle$ are rovibrational wavefunctions of $X$ and $a$ states, while $d^{ab}_{AX}(R)$ and $d^{ab}_{ba}(R)$ are the \emph{ab initio} TDM functions for the corresponding spin-allowed $A^1\Sigma^+-X^1\Sigma^+$ and $b^3\Pi-a^3\Sigma^+$ transitions, respectively.

\subsection{Fine and hyperfine structure of the $A\sim b\to a$ transitions}\label{Asen-hfs}

Intensity differences observed in P,R - doublets of the ``triplet'' $A\sim b\to a$ progressions can be explained in the framework of the Hund's (\textbf{b}) coupling case of the $a^3\Sigma^+$ state, which rotational levels consist of almost degenerate three $F_{1,2,3}$-components corresponding to the same quantum number $N_a = J_a, J_a\pm 1$~\cite{Bernath2005}. Indeed, both $P(N_a=J^{\prime}-1)$ and $R(N_a=J^{\prime}+1)$ branches of the $A\sim b\to a$ transition are compositions of the two overlapped $P^P(J_a=J^{\prime}+1)+P^Q(J_a=J^{\prime})$ and $R^R(J_a=J^{\prime}-1)+R^Q(J_a=J^{\prime})$ lines, respectively~\cite{Kato1993}, hence, the experimental LIF intensities should be defined by means of the double sum:
\begin{eqnarray}\label{Iteninterfer}
I^{P/R}_{A\sim b\to a} \sim \sum_{J_a}\left|\sum_{\Omega^{\prime}=0^+,1,2}\langle \phi_{b\Omega^{\prime}}^{J^{\prime}}|d^{ab}_{ba}|\chi_a^{N_a}\rangle
\frac{S_{\Omega^{\prime}J^{\prime}}^{N_aJ_a}}{3(2J^{\prime}+1)}\right|^2,
\end{eqnarray}
where the properly normalized rotational factors $S_{\Omega^{\prime}J^{\prime}}^{N_aJ_a}$ are the direction cosine matrix elements being the analytical functions of the rotational quantum numbers and their projections~\cite{Bernath2005, Field2004}.

Using the $P_{b2}\approx 0$ condition the general relation (\ref{Iteninterfer}) can be refined as
\begin{eqnarray}\label{IQ}
I_{P^P} &\sim& (J^{\prime}+2)|\mu_{0a}|^2;\quad I_{R^R} \sim (J^{\prime}-1)|\mu_{0a}|^2,\\
I_{P^Q} &\sim& J^{\prime}\left[\mu_{0a}-\sqrt{2}\mu_{1a}\sqrt{(J^{\prime}+1)/J^{\prime}}\right ]^2,\nonumber\\
I_{R^Q} &\sim& (J^{\prime}+1)\left[\mu_{0a}+\sqrt{2}\mu_{1a}\sqrt{J^{\prime}/(J^{\prime}+1)}\right ]^2,\nonumber
\end{eqnarray}
where
\begin{eqnarray}\label{overlap}
\mu_{0a}=\langle \phi_{b0^-}^{J^{\prime}}|d^{ab}_{b0^+a1}|\chi_a^{N_a}\rangle;\quad
\mu_{1a}=\langle \phi_{b1}^{J^{\prime}}|d^{ab}_{b1a0^-}|\chi_a^{N_a}\rangle
\end{eqnarray}
are the vibronic transition matrix elements between upper and lower electronic states corresponding to Hund's (\textbf{c}) coupling case. It should be noted that the matrix elements (\ref{overlap}) can be a bit different for the $P$ and $R$ lines since the rovibrational wavefunctions $\chi_a^{N_a}(R)$ are distinguished for the $N_a=J^{\prime}\pm 1$ levels. Hereafter, $d^{ab}_{b0^+a1}(R)$ and $d^{ab}_{b1a0^-}(R)$ are the so called ``perpendicular'' TDMs between different $\Omega$-components of $a$ and $b$ triplet states. These functions are slightly differing within the full relativistic approximation but they should stay identical within scalar-relativistic and non-relativistic approximations (see Sec.~\ref{zay-relativism} and Fig.~\ref{Fig_TDM_RbCs}).

Then, the additional assumption $J^{\prime}\gg 1$ straightforwardly transforms equations (\ref{IQ}) to a simple form:
\begin{eqnarray}\label{sum_ratio}
\frac{I_P}{I_R}=\frac{I_{P^P}+I_{P^Q}}{I_{R^R}+I_{R^Q}}\approx \left(\frac{\nu_P}{\nu_R}\right)^4\left|\frac{\mu^2_{0a}+[\mu_{0a}-\sqrt{2}\mu_{1a}]^2}{\mu^2_{0a}+[\mu_{0a}+\sqrt{2}\mu_{1a}]^2}\right |,~~
\end{eqnarray}
where the non-vanishing transition matrix element $\mu_{1a}$ can be responsible for the non-equal intensities of the $P$- and $R$-lines belonging to the same $v_a$-doublet. Note that the relation $[I_P+I_R]\sim [\mu^2_{0a}+\mu_{1a}^2]$ remains valid for all bands.

The HFS of the $a^3\Sigma^+$ state is determined by the Fermi contact interaction in Rb and Cs atoms being very similar to that observed in other alkali metal dimers~\cite{Kasahara1996} (see e.g.~\cite{Docenko2011, Pashov2005, Staanum2007}). Given the nuclear spin of $^{133}$Cs ($I=7/2$), $^{85}$Rb ($I=5/2$) and $^{87}$Rb ($I=3/2$) it is straightforward to calculate the HFS for both the $^{85}$Rb$^{133}$Cs and $^{87}$Rb$^{133}$Cs isotopomers using the approach from Ref.~\cite{Kasahara1996} and assuming the Hund's ($b_{\beta S}$) coupling case. In this case, first, the interaction between the electron spin and the spins of the two nuclei is accounted for and, second, the resultant intermediate angular momentum is coupled to the rotation of the nuclei ($N$)~\cite{Kasahara1996}. The HFS of the $a^3\Sigma^{+}$ state levels is independent of the rotational and the vibrational quantum numbers $N_a$ and $v_a$ except for the levels close to the asymptote, where the HFS interaction with the quasi-degenerate singlet $X^1\Sigma^{+}$ state takes place.

In Fig.~\ref{Fig_HFS} one can see a typical HFS of lines to the $a^3\Sigma^{+}$ state in $^{85}$Rb$^{133}$Cs and $^{87}$Rb$^{133}$Cs isotopomers. The calculated hyperfine splitting of the $a^3\Sigma^{+}$ levels is shown with bars below the line profiles. For the $^{87}$Rb$^{133}$Cs isotopomer the HFS consists of 96 components, while for $^{85}$Rb$^{133}$Cs the number of components is 144, some of them nearly degenerate. For these calculations the following values of atomic Fermi contact interaction constants were used: a$_F$($^{133}$Cs)=0.077, a$_F$($^{85}$Rb)=0.0337, and a$_F$($^{87}$Rb)=0.1139 cm$^{-1}$~\cite{Radzig1985}.

\subsection{The \emph{ab initio} $A/b - X/a$ transition dipole moments}\label{zay-relativism}

\subsubsection{Scalar-relativistic calculations}\label{CICPP}

The scalar-relativistic \emph{ab initio} transition dipole moments $d^{s-r}_{ij}(R)= | \langle \psi_i|\textbf{d}|\psi_j\rangle |$ for the spin-allowed $A^1\Sigma^+-X^1\Sigma^+$ and $b^3\Pi-a^3\Sigma^+$ electronic transitions in K$_2$, KCs and RbCs molecules were evaluated in the wide range of internuclear distances $R$. The electronic wave functions $\psi_j$ corresponding to a pure Hund's (\textbf{a}) coupling case were obtained within the framework of configuration interaction (CI) calculations combined with the approximate treatment of core-valence interaction by means of core polarization potentials~\cite{Fuentealba1983} (CPPs). The computational details can be found in our previous works~\cite{Kim2009, Pazyuk2016}. Briefly, the atomic cores of K, Rb, and Cs were replaced by the averaged relativistic effective core potentials from Ref.~\cite{Ross1990}, leaving nine electrons for explicit treatment. The optimized molecular orbitals were generated using the state-averaged complete active space self-consistent field (SA-CASSCF) method~\cite{Werner1985}, taking the (1-3)$^{1,3}\Sigma^+$, (1,2)$^{1,3}\Pi$, and (1)$^{1,3}\Delta$ electronic states with equal weights. The dynamic correlation was treated explicitly only for the two valence electrons within the multi-reference CI calculations~\cite{Knowles1992}. At both steps all sub-valence orbitals were kept doubly occupied. The rest core polarization and core-valence electron correlation effects were taken into account using the semi-empirical $l$-independent CPPs. The CI-CPP calculations were conducted using the MOLPRO package~\cite{MOLPRO2010}. The resulting  $d^{s-r}_{AX}(R)$ functions of the K$_2$, KCs, and RbCs molecules are presented in Figs.~\ref{Fig_TDM_K2KCs} and~\ref{Fig_TDM_RbCs}, respectively.

\subsubsection{Full relativistic calculations}\label{FSRCC}

Alternatively, the required \emph{ab initio} TDM functions for the K$_2$, KCs, and RbCs molecules were evaluated in the framework of fully relativistic electronic structure calculations corresponding to pure Hund's (\textbf{c}) coupling case.

For all alkali metal atoms the relativistic (two-component) semilocal shape-consistent effective core potentials (RECP) properly accounting for spin-dependent relativistic effects were used ~\cite{Mosyagin2010, GRECP}. Eight sub-valence (outer core) electrons, $(n-1)s^2\,(n-1)p^6$, and one $(nl^1)$ valence electron of each atom were described explicitly. Contracted Gaussian basis sets  $[7s\,7p\, 5d\, 3f\, 2g]$ for Rb and $[7s\,7p\,5d\,4f\, 3g\, 1h]$ for Cs are adopted from Ref.~\cite{Zaitsevskii2017}; a $[7s\,7p\,6d\,4f\,2g]$ basis set for K was built using the same principles.

To solve the relativistic 18-electron problem for all alkali diatomics treated we adopted the version of the Fock-space relativistic coupled-cluster (FSRCC) method~\cite{Eliav1998, Visscher2001} described in Ref.~\cite{Zaitsevskii2018a}. For all molecules the Fermi vacuum was defined by the ground-state SCF determinant for the doubly charged molecular ion, so that the neutral states corresponded to the two-particle Fock space sector ($0h2p$). The cluster operator expansion comprised only single and double excitations (FS-RCCSD approximation). The model spaces for the K$_2$, RbCs, and KCs were spanned by all possible distributions of two valence electrons among 52, 38, and 61 Kramers pairs of ``active'' (lowest-energy) spinors, respectively.

To suppress instabilities caused by intruder states, we employed, in the case of K$_2$ and RbCs, the dynamic denominator shift technique combined with the extrapolation to the zero-shift limit using matrix Pad\'e approximants. The details of the computational scheme are given in Appendix B. The calculations on K$_2$ and RbCs were performed with the appropriately modified DIRAC17 program package~\cite{DIRAC17,Saue2020}. In the case of KCs we used an alternative intruder state avoidance technique, based on the simulation of dynamic imaginary denominator shifts and implemented in the extensible tensor contraction code EXP-T~\cite{Oleynichenko2020}.

TDM matrix elements $d^{rel}_{ij}(R)=| \left<\psi_i\vert \textbf{d} \vert \psi_j\right> |$ between the adiabatic states with the relativistic wavefunctions $\psi_i$ and $\psi_j$ were evaluated in the frame of the finite-field (FF) scheme~\cite{Zaitsevskii2018} through the use of the finite-difference estimates for the derivative matrix elements in the r.h.s. of the following approximate relation:
\begin{equation}
\left<\psi_i\vert d_\eta\vert \psi_j\right>\approx\!\left(E_{j}-E_{i}\right)
\left<\tilde{\psi}^{\perp\perp}_j(F_\eta)\left|
\frac{\partial}{\partial F_\eta}\tilde{\psi}_i(F_\eta)\right.\right>
\left|\begin{array}{l}\\_{\!F=0}\end{array}\right.\!\! .
\label{fftdm}
\end{equation}
Here, $F$ is the applied uniform electric field strength, $\eta=x,\,y,\,z$, $\tilde{\psi}^{\perp\perp}$ and $\tilde{\psi}$ denote left and right eigenvectors of the field-dependent
non-Hermitian FSRCC effective Hamiltonian acting in the field-independent (constructed for $F=0$) model space, and $E_{j}$ and $E_{i}$ are the field-free energies for the states $j$ and $i$.
Although the calculations involved only the effective Hamiltonian eigenvectors (the model space projections of many-electron wavefunctions), the resulting transition moment value implicitly
incorporated the main contributions from the remainder (``outer-space'') parts of these wavefunctions~\cite{Zaitsevskii1998, Zaitsevskii2018}.

The approximate transition moments $\left<\psi_i\vert d_\eta \vert \psi_j\right>$ and $\left<\psi_j\vert d_\eta \vert \psi_i\right>^*$ defined by Eq.~(\ref{fftdm}) generally do not coincide. The nonphysical differences between these quantities obtained in our calculations were quite small (normally within 1\%) but not negligible. Therefore the final estimates were obtained by (\emph{a}) forcing the Hermiticity of the dipole moment matrix: $$\left<\psi_i\vert d_\eta \vert \psi_j\right> \longrightarrow\left(\left<\psi_i\vert d_\eta \vert \psi_j\right>+\left<\psi_j\vert d_\eta \vert \psi_i\right>^*\right)/2,$$ or (\emph{b}) the preliminary transformation of the non-Hermitian Bloch-type effective Hamiltonian to its Hermitian analog \emph{via} symmetric orthogonalization of its eigenvectors. These two approaches lead to practically identical results (differing only by ca. 10$^{-4}$ a.u.).

\begin{figure}%[t!]
\center
\includegraphics[scale=0.4]{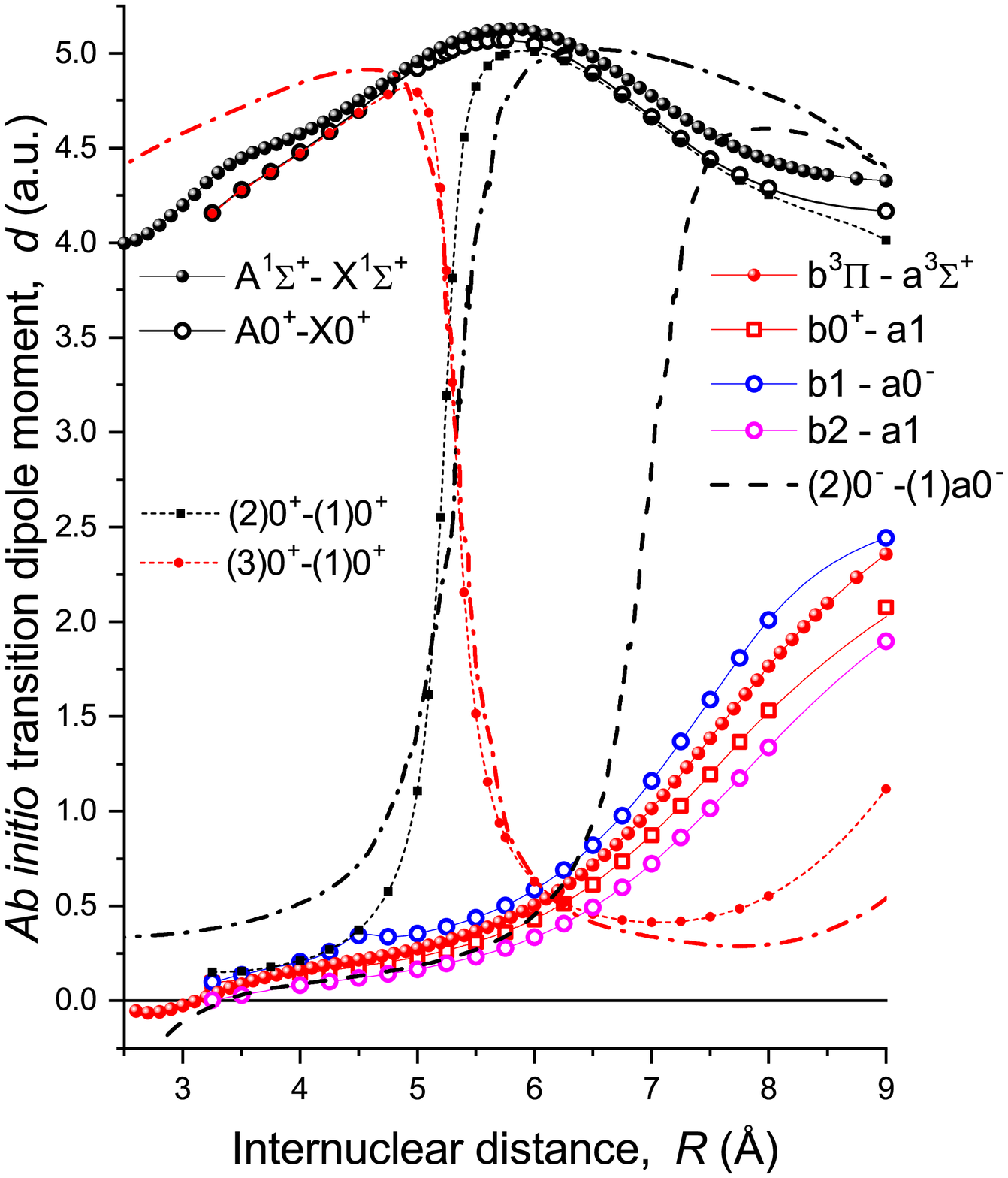}
\caption{The spin-allowed $A^1\Sigma^+-X^1\Sigma^+$ and $b^3\Pi-a^3\Sigma^+$ TDMs $d^{s-r}_{ij}(R)$ of the RbCs molecule obtained by the scalar-relativistic CI-CPP calculations. Their quasi-diabatic $A0^+-X0^+$, $b0^+-a1$, $b1-a0^-$ and $b2-a1$ analogues $d^{rel}_{ij}(R)$ were derived from the  full relativistic FSRCC-FF data. The fully relativistic $(2,3)0^+-(1)0^+$ curves are given as an example of the original TDM functions obtained using the conventional adiabatic representation. Dash-dot and dashed lines denote the relativistic $(2,3)0^+-(1)0^+$ and $(2)0^--(1)0^-$ TDM curves borrowed from Ref.~ \cite{Kotochigova2005}.}\label{Fig_TDM_RbCs}
\end{figure}

To cast the result of the FSRCC-FF calculations into quasi-diabatic form consistent with that used in the present CC deperturbation analysis (quasi-diabatic interatomic potentials, effective SO coupling matrix elements, and TDMs between quasi-diabatic states), we employed the technique based on projecting the solutions of scalar-relativistic eigenstates on the subspace of strongly coupled eigenstates of the full relativistic Hamiltonian~\cite{Zaitsevskii2017}. At this stage the many-electron wavefunctions were replaced by their FSRCC model space parts, i.~e. by the eigenvectors of the effective Hamiltonians. To transform the original adiabatic $(2,3)0^+-(1)0^+/(1)1$ TDM functions, we used the simple scheme based on the assumption that the lowest $X(1)0^+$ and $a(1)0^-/1$ states are almost pure spin states (singlet and triplet, respectively). In this case we determined the quasi-diabatic $A0^+-X0^+$ and $b0^+-a1$ TDMs by the requirement of vanishing the formally spin-forbidden $A0^+ - a1$ and $b0^+ - X0^+$ transitions. The resulting $d^{rel}_{ij}(R)$ functions relevant to the present work are presented in Figs.~\ref{Fig_TDM_K2KCs} and~\ref{Fig_TDM_RbCs}.

\section{Results and Discussion}\label{results}

\begin{table*}
\caption{Summary of experimental and theoretical results. Here, $\nu_L$ is the laser frequency used to excite the $A\sim b$ levels (in cm$^{-1}$). Symbol $^{\dag}$ marks the target levels of the $A\sim b$ complex. $E_{A\sim b}^{Expt}$ are experimental energies of the $A\sim b$ complex (in cm$^{-1}$) with the rotational quantum number $J^{\prime}$. $P_i$ are the calculated (in \%) fractional partitions of the $A\sim b$ complex, where $i\in[A^1\Sigma^+; b^3\Pi_{0^+}; b^3\Pi_1]$. $\tau^{Calc}_{A\sim b}$ (in ns) is the calculated radiative lifetime of the $A\sim b$ complex. $\mathbb{R}^{Expt}_{a/X}$ and $\mathbb{R}^{Calc}_{a/X}$ (in $10^{-4}$) are the experimental and theoretical branching ratios, respectively, obtained for the intercombination $A\sim b\to a/X$ transitions of the $^{85}$Rb$^{133}$Cs isotopomer ($^{\ddag}$ marks the $^{87}$Rb$^{133}$Cs isotopomer). $S^{FCF}_{ba}$ is the sum of FCF over the bound part of the $A\sim b\to a$ spectra. $v_a^{max}$ is the last quasi-bound vibrational level of the $a^3\Sigma^+$ state. $\mathbb{S}_{a/X}$ (in $10^{-3}$) is the sum of normalized to a particular $\nu_a$ intensities of the ``triplet'' $A\sim b\to a$ bands, which were evaluated according to Eq.(\ref{Intrestrict}) for the restricted number $\overline{v_a}$ of vibrational levels of the $a$-state. The abbreviation RCC$|$CPP denotes the results obtained using the present full-relativistic (FSRCC-FF) and scalar-relativistic (CI-CPP) TDM functions, respectively.}\label{Tab_Summary}
\begin{tabular}{cccccccccccccc}
%\begin{tabular}{llllllllllllll}
\hline\hline
$\nu_L$  & $J^{\prime}$ & $E^{Expt}_{A\sim b}$  & $P_A$ & $P_{b0^+}$ & $P_{b1}$ & $\tau_{A\sim b}^{Calc}$ & $\mathbb{R}_{a/X}^{Expt}$ & $\mathbb{R}_{a/X}^{Calc}$ & $S^{FCF}_{ba}$ & $v_a^{max}$ & $\mathbb{S}_{a/X}^{Expt}$ & $\mathbb{S}_{a/X}^{Calc}$ & $\overline{v_a}$\\
\hline
 &                &             &    &   &    & RCC$|$CPP &  & RCC$|$CPP & & & & RCC$|$CPP\\
\hline
10783.764  & $^{\dag}$60 & 10920.997 & 69.0 & 31.0 & 0.0 & 31.2$|$30.7 & 1.66 & 1.60$|$2.18 & 0.397 & 31 &  -  &      -      & -\\
                   & 48 & 11001.749  & 46.5 & 53.0 & 0.5 & 44.5$|$43.6 &  -   & 3.14$|$4.28 & 0.292 & 32 & 5.1 & 4.39$|$6.00 & 5-7\\
\hline
11008.935  & $^{\dag}$95 & 11286.359  & 43.8 & 54.2 & 2.0 & 46.2$|$45.3 & 4.79 & 3.94$|$5.24 & 0.251 & 28 &   -  &      -      & -\\
                    & 86 & 11780.884  & 38.2 & 61.8 & 0.0 & 50.6$|$49.4 &  -   & 8.20$|$11.0 & 0.253 & 29 & 11.6 & 21.9$|$29.6 & 0-2\\
\hline
11007.865 & $^{\dag}$97  & 11291.671  & 45.7 & 53.4 & 0.9 & 44.0$|$43.2 & 4.21 & 3.74$|$5.04 & 0.259 & 28 &  -   &      -      & -\\
           & ${\ddag}$91  & 11505.475  & 62.6 & 37.4 & 0.0 & 34.2$|$33.4 &  -   & 3.14$|$4.25 & 0.319 & 28 &  6.6 & 4.25$|$5.68 & 3-4\\
                    & 80  & 11722.342  & 40.5 & 59.4 & 0.0 & 47.9$|$46.9 &  -  & 7.39$|$10.8 & 0.332 & 29 & 11.8 & 19.3$|$26.1 & 0-3\\
                     & 100 & 11202.879  & 67.8 & 32.1 & 0.1 & 31.9$|$31.3 &  -  & 1.86$|$2.56 & 0.332 & 27 &  1.7 & 2.23$|$3.05 & 6-7\\
\hline
10743.007 &  $^{\dag}$40  & 10793.833 & 72.5 & 27.5 & 0.0 & 30.2$|$29.7  & 0.89 & 1.30$|$1.80 & 0.437 & 33 & -   & -  & -\\
\hline
10843.555 &  $^{\dag}$40  & 10894.372  & 60.3 & 39.6 & 0.1 & 34.6$|$34.0 & 2.01 & 2.14$|$2.93 & 0.387 & 33 & -    & -  & -\\
                    & 125 & 11230.157  & 72.0 & 28.0 & 0.0 & 32.4$|$31.8 &  -  & 1.09$|$1.49 & 0.206 & 25 & 7.5  & 3.04$|$4.16 & 8-10\\
                    & 133 & 11638.421  & 39.9 & 59.9 & 0.2 & 50.3$|$49.2 &  -  & 4.90$|$6.61 & 0.196 & 22 & 15.6 & 17.2$|$23.3 & 4-6\\
                    & 121 & 11550.602  & 39.9 & 59.4 & 0.7 & 50.2$|$49.2 &  -  & 4.73$|$6.35 & 0.201 & 22 & 9.0  & 10.2$|$13.8 & 4-6\\
                    & 131 & 11676.889  & 37.8 & 61.8 & 0.4 & 53.5$|$52.5 &  -  & 5.42$|$7.29 & 0.185 & 22 & 17.0 & 17.9$|$24.2 & 3-5\\
\hline
\end{tabular}
\end{table*}

Both experimental and theoretical radiative properties obtained for the particular rovibronic levels of the $A\sim b$ complex of the RbCs molecule are summarized in Table~\ref{Tab_Summary}. The experimental branching ratios $\mathbb{R}_{a/X}$ for the intercombination $A\sim b\to a/X$ transitions have been obtained according to Eq.~(\ref{Ratio}) for 5 targeted levels since other detected LIF progressions to the triplet $a$-state appeared to be too weak and fragmentary. The intensity distribution in the ``triplet'' $A\sim b\to a(v_a)$ progressions coming from accidentally excited levels of the RbCs $A\sim b$ complex was measured only for few selected $v_a$-bands having the maximal intensity. So, for these upper levels, in order to compare with calculations, the branching ratios $\mathbb{S}_{a/X}$ were introduced, which were defined as the sum over the restricted number of the vibrational levels $\overline{v_a}$ of the $a$-state (see Table~\ref{Tab_Summary}):
\begin{eqnarray}\label{Intrestrict}
\mathbb{S}_{a/X}=\frac{\sum_{\overline{v_a}}\sum_{N_a}I_{A\sim b\to a}}{I_{A\sim b\to X}(v_X)},
\end{eqnarray}
where $I_{A\sim b\to X}(v_X)$ is the intensity of the corresponding ``singlet'' $A\sim b\to X(v_X)$ transition selected for the normalization of the ``triplet'' bands. Note that the same normalization transition was always used for experimental and theoretical $\mathbb{R}_{a/X}$ and $\mathbb{S}_{a/X}$ estimates in the progressions.

For all observed ``singlet'' $A\sim b\to X(v_X)$ transitions the sum over vibrational $v_X$-levels of the ground singlet $X^1\Sigma^+$ state in Eq.~(\ref{Ratio}) was found to be corresponding to the so-called ``full'' LIF progressions, which satisfy the conventional sum rule of the Franck-Condon factors (FCFs) for bound-bound vibronic transitions: $\sum_{v_X} |\langle \phi_A|\chi_X\rangle|^2/P_A\approx 1$. In turn, for the ``triplet'' $A\sim b\to a(v_a)$ transitions the similar sums over bound $v_a$-levels of the triplet $a$-state
\begin{eqnarray}\label{SumFCF}
S^{FCF}_{ba} = \sum_{v_a}^{v_a^{max}} |\langle \phi_{b0^+}|\chi_a\rangle|^2/P_{b0^+}
\end{eqnarray}
are essentially less than 1 since the main contribution to the ``triplet'' decay channel comes from the $A\sim b\to a$ bound-continuum transitions.

The theoretical branching and intensity ratios, as well as the radiative lifetimes in Table~\ref{Tab_Summary}, were obtained by means of the calculated $A-X$ and $b-a$ TDM functions. The \emph{ab initio} TDMs of the RbCs molecule presented in Fig.~\ref{Fig_TDM_RbCs} clearly demonstrate that the singlet-singlet function $d^{ab}_{AX}(R)$ is about 10-25 times larger than the triplet-triplet function $d^{ab}_{ba}(R)$ at the region $R\in [3.5, 6.0]$~\AA~where the corresponding experimental LIF intensities are relevant. Hence, the strength of the ``triplet'' $A\sim b\to a$ transitions is expected to be much less than their ``singlet'' $A\sim b\to X$ counterparts, as observed in the experiment. Since the branching ratios of the $A\sim b\to a/X$ transitions obey the inequality $\mathbb{R}\ll 1$, the radiative lifetime of the $A\sim b$ complex is mainly determined by the $A-X$ decay channel:
\begin{eqnarray}\label{tau}
\frac{1}{\tau_{A\sim b}}&\approx& \sum_{v_X}\nu^3_{A\sim b-X}|\langle \phi_A|d_{AX}|\chi_X\rangle|^2\\
&\thickapprox& \langle \phi_A|[U_A-U_X]^3[d_{AX}]^2|\phi_A\rangle\nonumber,
\end{eqnarray}
where the approximate sum rule~\cite{Tellinghuisen1984, Pazyuk1994} is used to avoid a tedious summation over vibrational levels (including continuum) of the ground $X$-state. The resulting lifetimes (see Table~\ref{Tab_Summary}) strongly depend on the fraction of the singlet $A$-state: $\tau_{A\sim b}\approx \tau_A/P_A$, where $\tau_A\approx 19\div 21$~ns is the mean lifetime of the deperturbed $A$-state, which can be compared with the atomic Cs(6$^2$P) value~\cite{Cstau}~30.4$\div$34.8~ns.

It is seen from Fig.~\ref{Fig_TDM_RbCs} that the spin-orbit-free RbCs $A0^+-X0^+$ TDM function agrees very well with the scalar-relativistic $A^1\Sigma^+-X^1\Sigma^+$ TDM function at intermediate internuclear distances. Hence, the $\tau$-values estimated by Eq.(\ref{tau}) using the FSRCC-FF TDM function are only 0.5-1.0~ns higher than those obtained by means of the CI-CPP function. It should be noted that the CI-CPP functions derived for the K$_2$, KCs, and RbCs molecules approaching to atomic limit systematically overestimate their atomic K(4$^2$P-4$^2$S) and Cs(6$^2$P-6$^2$S) counterparts~\cite{Cstau, Atom} by 4\% and 7\%, respectively. At the same time, the full-relativistic FSRCC-FF functions reproduce the corresponding atomic transition probabilities with the unprecedentedly small error of~1-2\%.

\begin{figure}%[t!]
\center
\includegraphics[scale=0.40]{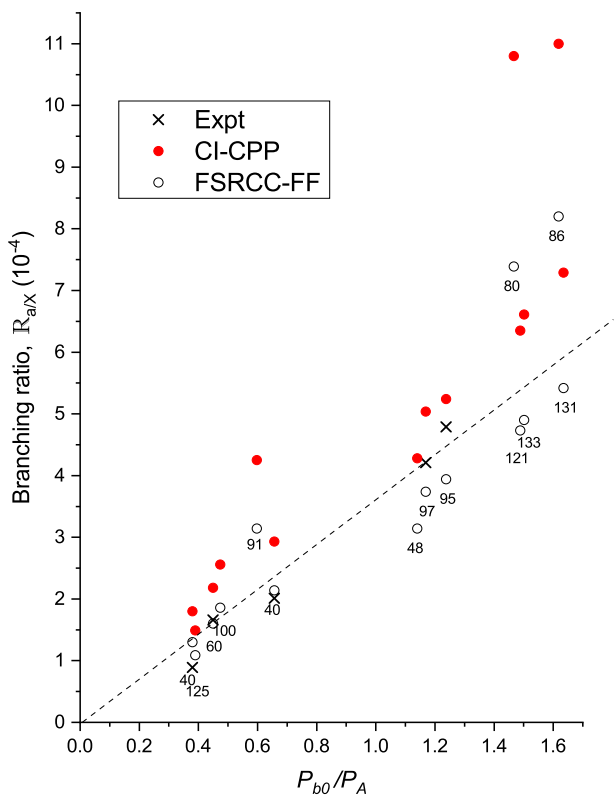}
\caption{Experimental and calculated branching ratios $\mathbb{R}_{a/X}$ obtained for the discrete parts of the $A\sim b\to a/X$ LIF progressions to triplet $a^3\Sigma^+$ and singlet $X^1\Sigma^+$ ground states as a function of the fractional mixing $P_{b0}/P_A$. The numbers below a symbol mark a rotational level $J^{\prime}$ of the $A\sim b$ complex.} \label{Fig_ratio}
\end{figure}

It is useful to highlight a general impact of the singlet-triplet mixing in the $A\sim b$ complex on the branching ratio of the $A\sim b\to a/X$ transitions in a graphical form. Indeed, Fig.~\ref{Fig_ratio} presents both experimental and calculated $\mathbb{R}_{a/X}$-values as a function of the ratio of the triplet and singlet fractions $P_{b_0}/P_A$ in the total multi-component wave function of the $A\sim b$ complex. The depicted $\mathbb{R}_{a/X}$-values clearly demonstrate the steep linear growth of the branching ratio with increasing the triplet component contribution. The exception is $\mathbb{R}_{a/X}^{calc}$ observed for the levels with $J^{\prime}$= 80, 86 ($^{85}$Rb$^{133}$Cs) and $J^{\prime}$= 91 ($^{87}$Rb$^{133}$Cs). The reason for this discrepancy is not clear at the moment. According to Fig.~\ref{Fig_ratio} and Table~\ref{Tab_Summary} the theoretical branching $\mathbb{R}_{a/X}^{calc}$ values agree well with the experiment. Note, that the uncertainty of $\mathbb{R}_{a/X}^{expt}$ values caused by several scaling factors in intensity measurements could be about 10-15\%. The $\mathbb{S}_{a/X}^{Calc}$ values agree satisfactorily as well with $\mathbb{S}_{a/X}^{Expt}$ values except the levels with $J'=80,86$. It is seen also that the CI-CPP $\mathbb{R}_{a/X}^{calc}$ values are systematically higher than the respective FSRCC-FF values. 

\begin{figure}%[t!]
\center
\includegraphics[scale=0.35]{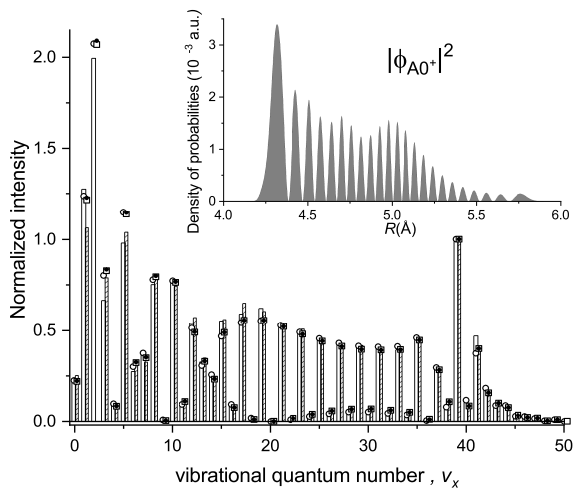}
\caption{Experimental and theoretical relative intensity distributions in the full ``singlet'' $A\sim b\to X^1\Sigma^+(v_X)$ LIF progression starting from the upper level $J^{\prime}$ = 97, $E_{A\sim b}$ = 11~291.672 ~cm$^{-1}$. All intensities are normalized with respect to the $v_X$ = 39 band. Empty and striped columns stand for experimental values of $R$- and $P$-branches, respectively. Points stand for theoretical values: empty circles and squares mark $R$- and $P$-branches, respectively, calculated within the FSRCC-FF approach. Black points mark the $P$-branch calculated within the CI-CPP approach. Note that all theoretical intensities for a particular band almost coincide. The inset shows the radial probability density distribution of the non-adiabatic vibrational wavefunction corresponding to the $A^1\Sigma^+$ state.} \label{Fig_intensity2}
\end{figure}

\begin{figure}%[t!]
\center
\includegraphics[scale=0.38]{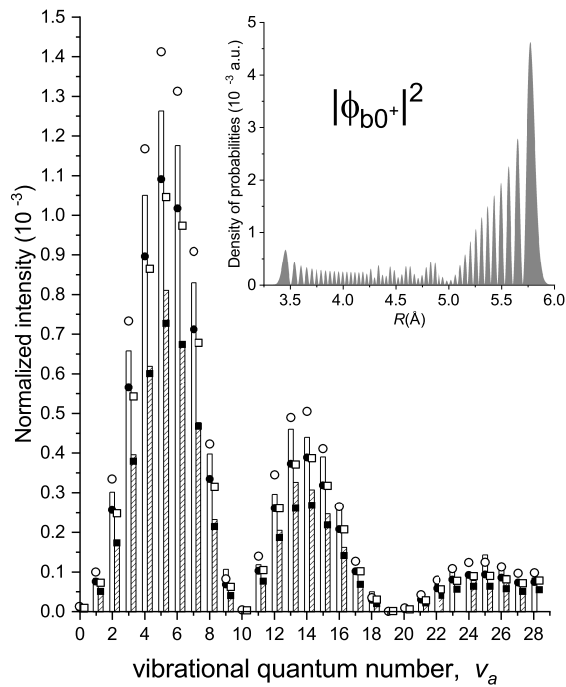}
\caption{Comparison of experimental and theoretical intensity distributions in the discrete part of the ``triplet'' $A\sim b\to a^3\Sigma^+(v_a)$ LIF progression starting from the upper level $J^{\prime}$ = 97, $E_{A\sim b}$ = 11~291.672 ~cm$^{-1}$. All intensities are normalized to the $v_X$ = 39 band of the ``singlet'' $A\sim b\to X(v_X)$ progression, see Fig.~\ref{Fig_intensity2}. Empty and striped columns are experimental values of the $R$- and $P$-branches, respectively. Points stand for theoretical values: full circles and squares mark $R$- and $P$-branches, respectively, calculated within the FSRCC-FF approach; empty circles and squares mark $R$- and $P$-branches, respectively, calculated within the CI-CPP approach. The inset shows the radial probability density distribution of the non-adiabatic wavefunction corresponding to the $b^3\Pi_{0^+}$-component.}
\label{Fig_intensity1}
\end{figure}

A relative intensity distribution over ground $X$-state vibrational levels $v_X$ for the ``singlet'' $A\sim b\to X^1\Sigma^+(v_X)$ LIF progression starting from the $J^{\prime}$= 97 level of the $A\sim b$ complex is presented in Fig.~\ref{Fig_intensity2}. The distribution is normalized with respect to the transition to the $v_X$ = 39 band since it falls within the spectral range around 9~300~cm$^{-1}$, for which the spectral sensitivity curve $S(\nu)$, is more reliable than for high frequency range, as follows from Fig.~\ref{Fig_sensitivity}. A comparison of normalized experimental and theoretical intensity distributions demonstrates overall excellent agreement for both theoretical approaches. Fig.~\ref{Fig_intensity1} presents a relative intensity distribution into the ``triplet'' $A\sim b\to a(v_a)$ LIF progressions from the same upper level $J^{\prime}$ = 97 over vibrational levels $v_a$ of the $a^3\Sigma$ state. Here, the same normalization as in Fig.~\ref{Fig_intensity2} is preserved. One can see that the overall LIF intensity is about 3-4 orders of magnitude smaller. Though the theoretical relative intensity distribution agrees very well with the experimental one, the values of $I^{Calc}_{A\sim b\to a} (v_a)$ normalized to $v_X$ = 39 of ``singlet'' progressions systematically slightly diverge from the experimental intensities in the ``triplet'' progression. Indeed, while the FSRCC-FF values are systematically lower, the respective CI-CPP values are higher than the experimental counterparts. This difference apparently raises from the systematic error inevitably taking place when using the extremely small (by absolute value) $b-a$ TDM functions. A similar trend was observed also for  $A\sim b\to a(v_a)$ LIF progression arising from the level with $J'=95$

\begin{figure}%[t!]
\center
\includegraphics[scale=0.34]{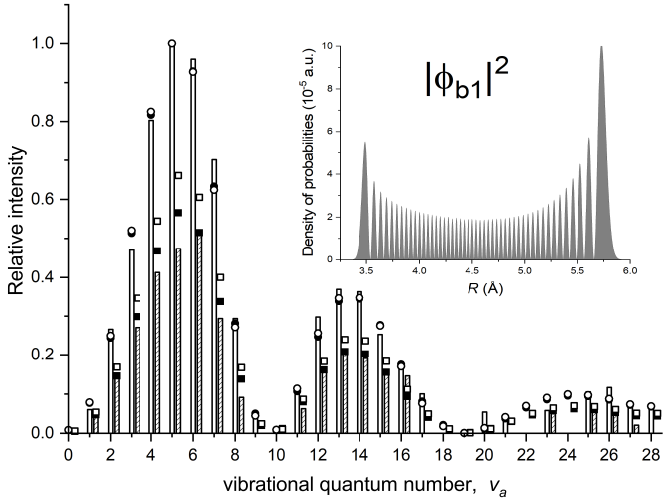}
\caption{Experimental and calculated intensity distributions of the $A\sim b\to a$ LIF progression starting from the upper level $J^{\prime}$ = 95, $E_{A\sim b}$ = 11~286.359~cm$^{-1}$. The same notations are used as in Fig.~\ref{Fig_intensity1}. All intensities are normalized to the most intensive $R$-line of the $v_a$ = 5 band. The density of the vibrational wavefunction corresponding to the $b^3\Pi_1$-component is shown in the inset.}\label{Fig_Inten_ratio95}
\end{figure}

Fig.~\ref{Fig_intensity1} demonstrates a marked intensity difference between $R$ and $P$-branches. This difference, which was observed in several ``triplet'' $A\sim b\to a(v_a)$ LIF progressions, can be attributed to the influence of the $b^3\Pi_1$-component of the $A\sim b$ complex. Expression (\ref{sum_ratio}) predicts a possible deviation of the $I_P/I_R$ ratio from 1 for the $A\sim b$ levels, in which the fraction of the $b^3\Pi_1$-component in the multi-channel CC vibrational eigenfunction is small but not negligible. Indeed, according to Fig.~\ref{Fig_intensity1} the expression (\ref{sum_ratio}) reproduces rather well the ``anomalies'' of the $I_P/I_R$ ratio for the $J^{\prime}$ = 97 level, which possesses only 0.9\% of the $b^3\Pi_1$-component, see Table~\ref{Tab_Summary}. Another example is presented in Fig.~\ref{Fig_Inten_ratio95}, which demonstrates a very good agreement between experimental and theoretical intensity distributions within the $R$- and $P$-branches of the $A\sim b\to a(v_a)$ progression starting from the $J^{\prime}$ = 95 level having a minor (2\%) admixture of the $b^3\Pi_1$-component. In both cases, the relativistic FSRCC-FF estimates based on the slightly differing $b0^+-a1$ and $b1-a0^-$ TDM functions (Fig.~\ref{Fig_TDM_RbCs}) are found to be closer to the experimental observation than the scalar-relativistic CI-CPP values calculated with a single $b^3\Pi-a^3\Sigma^+$ TDM function.

Radial probability density distributions depicted in the insets of Figs.~\ref{Fig_intensity2} and~\ref{Fig_intensity1} highlight an impact of the strong mutual perturbations on a nodal structure of the non-adiabatic vibrational $\phi_{A0^+}$ and $\phi_{b0^+}$ wavefunctions leading to dramatic changes of the corresponding FCF values as compared to their adiabatic counterparts~\cite{Pazyuk2019, Pupyshev2010}. At the same time, the shape of the weakly perturbed $\phi_{b1}$ wavefunction (see the inset in Fig.\ref{Fig_Inten_ratio95}) is still close to the conventional adiabatic form.

\section{Conclusions}

The relative error in the \emph{ab initio} TDMs obtained for the strong spin-allowed $A - X$ transition in RbCs, KCs, and K$_2$ molecules by means of scalar-relativistic and full-relativistic methods seems not to exceed a few percent in the region, where the corresponding experimental $A\sim b\to X(v_X)$ LIF intensities are relevant. Furthermore, the \emph{a priori} most advanced FSRCC-FF method provides the $A0^+-X0^+$ TDM function, which, at large internuclear distance, deviates by only 1-2\% from its atomic counterpart. This means that both presently used $A-X$ TDM functions describe almost perfectly the relative intensities predicted for the strong ``singlet'' $A\sim b\to X$ progressions, which, hence, can be used to calibrate the relative sensitivity of the FT spectrometer detection system in a wide spectral range. The TDMs calculated both in scalar and full relastivistic approaches are provided in numerical form in the supplementary materials \cite{Suplementary} 

At the same time, a reliability of the theoretical estimates obtained for very weak ``triplet'' $A\sim b\to a(v_a)$ progressions of the RbCs molecule is mainly determined by the accuracy of the $b - a$ TDM functions, which are extremely small at the internuclear distances covered by the present experiment. In these circumstances, even small experimental intensities (corresponding to the spontaneous emission Einstein's coefficients less than $10^4$~s${}^{-1}$) can be still useful to probe and validate their \emph{ab initio} counterparts.

The quasi-diabatic $d^{rel}_{AX}(R)$ function obtained by FSRCC-FF method predicts an absolute strength of the ``singlet'' $A\sim b\to X$ transitions and radiative lifetimes of the $A\sim b$ complex in the RbCs molecule with an uncertainty of about 2-5\%, while the corresponding ``triplet'' $A\sim b\to a$ transition probabilities can be estimated, by means of the corresponding $d^{rel}_{ba}(R)$ functions, with a systematic error within 10-50\%. Nevertheless, the absolute accuracy achieved for both $A\sim b\to a/X$ transitions looks sufficient for modeling radiative properties of the stimulated Raman adiabatic passage (STIRAP) process, $a\to A\sim b \to X$, at the required spectroscopic level of confidence.

\section{Acknowledgments}
Riga team acknowledges the support from the Latvian Science Council Grant LZP2018/5: ``Determination of structural and dynamic properties of alkali diatomic molecules for quantum technology applications'' and from the University of Latvia Base Funding No A5-AZ27; A.K. acknowledges the support from the Post-doctoral Grant No. 1.1.1.2/16/I/001, proposal No. 1.1.1.2/I/16/068. A.V.S. acknowledges the partial support from Russian Foundation for Basic Research, RFBR grant No.18-03-00726. A.P. acknowledges partial support from National Scientific Fund of Bulgaria through Grant DN18/12/2017.

\section{Appendix-A}
The rovibronic eigenvalues and eigenfunctions for the $A\sim b$ complex of K$_2$, KCs, and RbCs molecules were obtained by solving four CC equations:
\begin{eqnarray}\label{CC}
\left(- {\bf I}\frac{\hbar^2 d^2}{2\mu dR^2} + {\bf V}(R;\mu,J^{\prime}) - {\bf I}E^{CC}_{A\sim b}\right)\mathbf{\Phi}(R) = 0
\end{eqnarray}
with the conventional boundary $\phi_i(0)=\phi_i(\infty)=0$ and normalization $\sum_{i=1}P_i=1$ conditions, where $P_i=\langle\phi_i|\phi_i\rangle$ is the fractional partition of the non-adiabatic $A\sim b$ state and $i\in[A^1\Sigma^+;b^3\Pi_{0^+};b^3\Pi_1;b^3\Pi_2]$. The relevant potential energy matrix ${\bf V}(R;\mu,J^{\prime})$ was taken in the form:
\begin{eqnarray}\label{Ham}
\langle ^{1}\Sigma^{+}|H|^{1}\Sigma^{+} \rangle & = & U_A + B(X+2), \nonumber \\
\langle ^{3}\Pi_{0^+}|H|^{3}\Pi_{0^+} \rangle & = & U_{b0^+} + B(X+2),\nonumber \\
\langle ^{3}\Pi_{1}|H|^{3}\Pi_{1} \rangle & = & U_{b0^+} + A^{so}_{10} + B(X+2),\nonumber \\
\langle ^{3}\Pi_{2}|H|^{3}\Pi_{2} \rangle & = & U_{b0^+} + A^{so}_{10} + A^{so}_{12} + B(X-2),\nonumber \\
\langle ^{1}\Sigma^{+}|H|^{3}\Pi_{0^+} \rangle & = & - \sqrt{2}\xi^{so}_{Ab0}, \\
\langle ^{3}\Pi_{0}|H|^{3}\Pi_{1} \rangle & = & - B\sqrt{2X},\nonumber \\
\langle ^{3}\Pi_{1}|H|^{3}\Pi_{2} \rangle & = & - B\sqrt{2(X-2)},\nonumber\\
\langle ^{1}\Sigma^{+}|H|^{3}\Pi_{1} \rangle & = & - \eta B\sqrt{2X}, \nonumber
\end{eqnarray}
where $X = J(J+1)$, $B = \hbar^{2}/2 \mu R^{2}$ while $U_A(R)$, $U_{b0^+}(R)$ are the effective (quasi-diabatic) interanucler potentials of $A^{1}\Sigma^{+}$ and $b^{3}\Pi_{0^+}$ states. $\xi^{so}_{Ab0}(R)$ is the off-diagonal SO coupling function while $A^{so}_{10}(R)$ and $A^{so}_{12}(R)$ are the diagonal $\Omega=1 - 0^{+}$ and $\Omega = 2 - 1$ SO splitting functions of the triplet $b$-state, respectively. Numerical recipes implemented to solve the present CC problem can be found in Ref.~\cite{Yurchenko2016}.

\section{Appendix-B}
A straightforward application of the conventional FSRCC method in wide ranges of internuclear separations would inevitably lead to numerical instabilities due to the so-called intruder state problem ~\cite{Evangelisti1987,Zaitsevskii2017}, which manifests itself as the appearance of zero or small energy denominators in the FSRCC amplitude equations at certain nuclear geometries. Following Refs.~\cite{Zaitsevskii2017, Zaitsevskii2018a}, we suppressed these instabilities via replacing the conventional FSRCC energy denominators \cite{Eliav1998, Visscher2001} $D_K$, where $K$ stands for an excitation from the model space to its orthogonal complement (outer space), by their ``dynamically shifted'' counterparts
\begin{eqnarray}\label{shift} \nonumber
\\ D_K\;\longrightarrow \;D_K^\prime=D_K+S_K\,
\left(\displaystyle \frac{S_K}{D_K+S_K}\right)^n,
\; n\ge 1.~~~
\end{eqnarray}
The shift parameters $S_K$ (normally universal for all classes of excitations) should be chosen to ensure negative (and not small) $D_K+S_K$ values for all excitations $K$. In the present calculations all $S_K$ values were determined by the only one parameter $S_2$. For double excitations involving two valence particles we assumed $S_K = S_2$, whereas for the excitations affecting one valence particle we used $S_K = S_2/2$. Calculations for the Fermi vacuum sector were always performed with conventional non-shifted energy denominators. The denominator shifts are damped for well-defined (large negative) original energy denominators so that the errors in the computed properties of low-lying electronic states arising from the substitution (\ref{shift}) should decrease with the increase of $n$.At the same time, the use of large $n$ values destroys the convergence of iterative procedures of solving the FSRCC equations. To bypass this difficulty we performed a series of calculations with rather small consecutive $n$-values ($n$ = 2, 3, 4) followed by the [0/1] Pade extrapolation of the resulting FSRCC effective Hamiltonians to the infinite-$n$ (i. e. zero-distortion) limit~\cite{Zaitsevskii2018a}. To improve the stability of the extrapolation procedure, all effective Hamiltonians were pre-projected onto the subspace spanned by only those eigenstates of the effective Hamiltonians evaluated with $n = 4$ converging to the 3 lowest dissociation limits. Transition energies and transition dipoles derived from the extrapolated effective Hamiltonians exhibited only a very weak dependence on the choice of the parameter $S_2$; the results presented in the paper were obtained with $S_2 = -0.4$ (K$_2$), -0.6 (KCs) and -0.4 (RbCs) a.u. Furthermore, the TDM estimates calculated with $n=3,\,4$ practically coincided with those obtained with the help of the extrapolation procedure. This enabled us to use single-calculation estimates for the TDM function of the KCs molecule.

\bibliography{rbcs.bib}

\end{document}